 \documentclass[ 12pt]{article} %NOT BIBER
\usepackage[utf8]{inputenc}
\usepackage[backend=biber]{biblatex}
\usepackage{amsmath,amssymb}
\usepackage{xcolor,graphicx}       % professional-quality tables
\usepackage{doi}

\title{Parameterized sequential functions for temporal properties}
\author{ \href{https://orcid.org/0000-0001-5085-9794} {\includegraphics[scale=0.06]{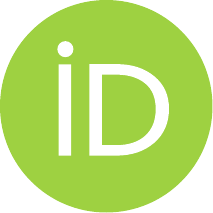}\hspace{1mm}Victor Yodaiken}\thanks{
		Independent Researcher.} \\
	Austin Texas\\
	\texttt{victor.yodaiken@gmail.com} \\
}
\date{October 2025}
\newcommand{\ess}{\epsilon}
\newcommand{\xy}{\cdot}
\newcommand{\concat}{\mathop{\mathtt{concat}}}

\newcommand{\always}{\mathop{\mathit{Always}}}
\newcommand{\until}{\mathop{\mathit{Until}}}
\newcommand{\eventually}{\mathop{\mathit{Eventually}}}

\newcommand{\next}{\mathop{\mathit{Next}}}
\newcommand{\seq}[1]{\langle #1\rangle}
\newcommand{\set}[1]{\{ #1\}}
\newtheorem{defn}{Definition}
\newtheorem{spec}{Specification}
\newtheorem{thm}{Theorem}
\newcommand{\beq}{\begin{equation}}
\newcommand{\eeq}{\end{equation}}
\begin{document}

\maketitle
\begin{abstract}
	Temporal logic is based on an intuitively appealing claim
	that being able to specify \emph{when} a property will be true
	can be useful in specifications of computer software and 
	hardware - particularly in \emph{reactive} systems
\cite{HarelReactive}. This paper shows that
	temporal properties are among the properties that 
	can be expressed
	directly and more simply in terms of state machines using
	sequential functions\cite{yodaikenlarge} to define state
machine behavior and compositional architecture. Specifications and
proofs can be carried out in 
ordinary algebra without the need for a formal
language or axiomatic proof system. The compositional properties of 
sequential functions also simplify specification of composite and 
concurrent systems. 
\end{abstract}
\section{Introduction}

\begin{quote}\emph{
Everybody who has worked in formal logic will confirm that it is one of the technically most refractory parts of mathematics.} --- Von Neumann\cite{vonneumann} \end{quote}

Temporal logic\cite{krithikeller,Pnueli, Manna,Clarke,karp,LamportTLA} is based on an
 intuitively appealing claim that being able
to specify \emph{when} a property must be true can be useful
in specifications of computer systems. Examples include properties like:
``\emph{Always} no more than one process can be accessing 
a mutually exclusive resource'',
and ``\emph{Eventually} every scheduled process must start to run'', and 
``a server cannot send an acknowledgment \emph{Until} it has received the request.''
This paper shows how to express similar 
and more detailed properties without formal logics
or other axiomatic system using  
\emph{sequential functions} \cite{yodaikenlarge}. A sequential
function is a map from finite sequences of events to outputs so
that 
\(f(w)\) is the output of a discrete state system
in the state
reached by following event sequence \(w\) from the initial 
state. As seen below, these functions 
can specify detailed operation of large scale state
systems as well as abstract properties
and the behavior and architecture of composite and concurrent systems.

Section \ref{sec:basics} starts with 
a short introduction to sequential functions and 
a definition
of temporal-logic-style operators on boolean valued
sequential functions so that 
\((\always P)(w)\), \((\eventually P)(w)\) and so on are well-defined
and can be seen to be at least similar to the corresponding
temporal qualifiers (which differ among themselves in any case).
Section \ref{sec:examples} tackles a number of examples, including
real-time and concurrent systems, and contrasts specifications that do and do not use the temporal-logic-like operators. 
Section \ref{sec:abadi} covers a real-time example
used in a paper by Abadi and Lamport\cite{abadilamport} but unlike
that paper includes a complete proof. The final 
section tries for some context.

The goal of this project is perhaps anachronistic:
it is to provide human developers of operating systems and
other complex ``reactive'' software \cite{HarelReactive} with a
mathematical basis for specifying and thinking about designs. The
reader can form their own opinion about how successful
this effort has been.

\section{Basic methods and Temporal-like operators}\label{sec:basics}

Each sequential function is associated with an \emph{alphabet} -- 
the set of discrete symbols representing 
the events that can can produce a state change. 
State systems are deterministic, so any finite sequence
of events determines the state reached by following the sequence
from the initial state.
Event sequences in this paper are always finite. Much of the expressive
power of sequential functions derives from how convenient it is to 
make partial specifications: instead of making nondeterminism an
intrinsic property, we can just leave some behavior unspecified. 

Following the zero length sequence \(\ess\) from the initial state
leaves the system state unchanged
so for sequential function \(f\),  \(f(\ess)\)
is the output in the 
initial state. Appending an event \(a\)  to a sequence, \(w\xy a\), 
drives the system to the next state so \(f(w\xy a)\) is the output
in the state reached \emph{after} event \(a\)
drives the system one event from the state determined by \(w\). 

If \(z\) is a finite sequence, then \(w\concat z\)
concatenates \(z\) to \(w\) on the right and \(f(w\concat z)\) is the output
in the state reached by following \(z\) from the state determined by \(w\).
Primitive recursion on sequences completely defines a sequential function:
\[ 
\text{ for constant }c,\ f(\ess)= c,\text{ and } f(w\xy a) = g(a,f(w)).\]

Simple composition \[f_2(w) = h(f_1(w))\]
modifies the output. Compositions of
multiple sequential functions e.g. \[f(w) = (f_1(w),\dots f_n(w))\] define direct products of state systems (which do not interact).
An analog of simultaneous recursion of arithmetic primitive recursive
functions \cite{peter} defines a system of interacting components:
\[f(w) = (f_1(w_1),\dots f_n(w_n))\]
where each \(w_i = w_i(w)\) is a primitive recursive sequential function
that determines component interactions. 
See section \ref{sec:multiple} for details.

Boolean valued sequential functions 
that have values in the set \(\set{0,1}\) 
with \(1\) for true and \(0\) for false can be used to specify properties.
Boolean expressions like \(f(w) < c\) will be treated as abbreviations for characteristic boolean functions like
\[P(w) = \begin{cases} 1&\text{if }f(w)< c\\
	 0&\text{otherwise}\end{cases}\]

Sequential functions depend on a single event sequence argument, which
is usually given as the
first argument. Multiple arguments
can mean either that 
that the output set \(X\) consists of maps (generally
not sequential functions), so \(f(w,x,y,z)= (f(w))(x,y,z)\).

The focus here is on intuition and applications not the 
underlying automata theory. 

\subsection{Temporal style operators}\label{sec:temporal}

\begin{defn}
\[(\always P)(w) = 
		\forall z\text{ where } P(w\concat z)\text{ is defined,  }P(w\concat z)\]
\[(\eventually P)(w) = 
		\begin{array}[t]{l}
			\exists n\geq 0\text{ so that }\\
			\forall z\text{ where } P(w\concat z)\text{ is defined}
			\text{  and }length(z)\geq n,\\ P(w\concat z')
			\text{ for some prefix }z'\text{ of }z
		\end{array}\]
 \[(P\until Q)(w) = 
	\forall z\text{ where } P(w\concat z)\text{ is defined  and } P(w\concat z)=0\]
		\[Q(w\concat z')\text{ for some prefix }z'\text{ of }z\]

	\[(\next P)(w)= \begin{array}[t]{l}\text{for every }a\text{ in the alphabet of }P\\
		\text{ if } P(w\xy a) \text{ is defined } P(w\xy a)\\
	\end{array}\]
\end{defn}
Sequence prefixes and sequence length are well known operations,
but definitions can be found
in appendix \ref{appendix:sequences} in case of doubt.

Note that if \(P(w)\) then \((\eventually P)(w)\) with \(n=0\).
`
	The treatment of undefined states above is not necessarily the best for every application. It might also to be good to require that neither next nor eventually 
	are trivially true, with at least one \(a\) so that \(P(w\xy a)\) is
	defined and
	one sufficiently long sequence \(z\) so that \(P(w\concat z)\) is defined.

\subsubsection{Some Conventions}

	An expression \((\eventually P(w,x))\) has the meaning \((\eventually P_x)(w)\) where
	\(P_x(w) = P(w)(x)\)

Nested temporal style operators sometimes require a little care -- like 
nested summations or integrals do.
Consider \((\always (\eventually P) )(w)\).
Let \(Q(w) =  (\eventually P)(w)\). Then the original expression 
can be written \( (\always Q)(w)\) which means that for all \(z\) so that
\(Q(w\concat z)\) is defined, \(Q(w\concat z)\). 
Expanding \(Q\) the last expression means \[(\eventually P)(w\concat z)\] which means that there is some
\(n\) so that if \(u\) has length \(n\) or more, for some prefix \(u'\) of
\(u\), \[P(w\concat z \concat u').\]

Suppose
\(Sent(w,i,m)\) is true iff network
element \(i\) has sent message \(m\), \(Received(w,i,m)\) has
the obvious meaning and \(Sending(w,i,m)\) is true
if \(i\) is currently sending the message. For these, suppose 
that \(Sent(w,i,m)> Sent(w\xy a,i,m)\) is impossible
so that a message never
becomes unsent, and the same for \(Received\). Then
\begin{equation}\label{eq:sent}
	\text{If }Sending(w,i,m)\text{ then } (\eventually Sent)(w,i,m)
\end{equation}
would be a convenient property.  
If message \(m\) must be acknowledged with message \(m'\) and
	\begin{equation}\label{eq:eventuallyr}
		\begin{array}{l}
		\text{If } Sent(w,i,m) > Received(w,i,m')\\ \text{ then } (\eventually Sending)(w,i,m)) \end{array}
	\end{equation}
requires that element \(i\) keep sending the message until it gets the acknowledgment. 

Alternatively, define \((Since\ P)(w)\) to be the count of
events since \(P\) first became true.
\[(Since\ P)(\ess) = 0\]
\[(Since\ P)(w\xy a)= \begin{cases}
	P(w) &\text{if }(Since\ P)(w) = 0\\
1+(Since\ P)(w)&\text{otherwise}\end{cases}\]

Then
for some \(n\geq 0\), 
\( (Since\ Sending)(w,i,m))*(1-Sent(w,i,m)))< n\) 
is equivalent to \ref{eq:sent} above and 
for some \(n_1\geq 0\) \( ((Since Sent)(w,i,m))*(1- Received(w,i,m')) < n_1\)
for some fixed \(n_1\), is the same constraint as in 
\ref{eq:eventuallyr} above. These constraints assume that there
are no errors, which is not always a good assumption. 
It may be better to allow the network element to conclude that
a client has failed or is unreachable if for some \(k\) 
\((Since\ Sent)(w,i,m)> k\) and \(Received(w,i,m)=0\).
Section \ref{sec:realtime} shows how to count elapsed time instead
of the number of events. 

\section{Examples}\label{sec:examples}

\subsubsection{Simple examples}\label{sec:simple}
Some simple examples illustrate and will be useful in section \ref{sec:abadi}. The common theme is to define state recursively.
The functions ``\(tail\)''  and ``\(head\) used below
are as usual and are defined in appendix \ref{appendix:sequences} in case
of doubt.
\begin{spec}\label{spec:simple} Components.\\
\begin{itemize}
	\item 	\(S\) is a \textbf{store} over values \(V\) if \(S(w\xy v)=v\) for
		every \(v \in V\).  It is a store over values \(V\) with initial
		value \(v_0\) if \(S(\ess)=v_0\). Otherwise the initial 
		value is not specified.
\item 	\(T\) is a \textbf{toggle} with initial value \(b\in \set{0,1}\)
	if \(T(\ess)=b\) and \(T(w\xy a) = 1- T(w)\).
\item 	\(Q\) is a \textbf{transparent queue} over set \(V\) if
	 \(Q(\ess)=\ess \) and
	\[Q(w\xy a) = \begin{cases} 
		Q(w)\xy v&\text{if }a= (enq,v)\\
		  tail(Q(w))&\text{if }a= deq\text{ and }Q\neq \ess\\
	Q(w)&\text{otherwise}\end{cases}
	\]
\item 
	\(Q\) is a  \textbf{bounded queue} over set \(V\) with length \(k\)
	if \(Q(\ess)=\ess\) and:
	\[Q(w\xy a) = \begin{cases} 
		Q\xy v&\text{if }a= (enq,v)\text{ and } Length(Q(w))<k\\
		  tail(Q) v&\text{if }a= deq\text{ and }Q\neq \ess\\
	Q(w)&\text{otherwise}\end{cases}\]
\item \(Q\) is a  \textbf{closed queue} over set \(V\) with length \(k\)
	if for some bounded queue with length \(k\), \(Q_{k}\) over \(V\)
	\[ Q(w)=\begin{cases}
		\ess&\text{if }Q_{k}(w)=\ess;;\\
	head(Q_{k}(w))&\text{otherwise}\end{cases}\]
		Here, \(\ess\not\in V\) is assumed.
\end{itemize}
\end{spec}

\subsection{Process scheduling}\label{sec:process}

\begin{figure}[h]
	\begin{center}
	\includegraphics[width=0.7\textwidth]{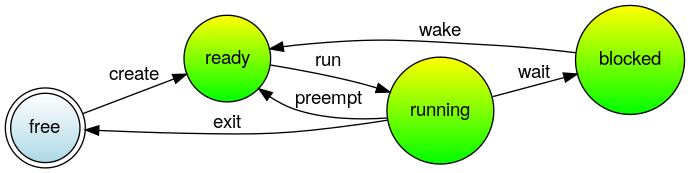}
	\end{center}
	\caption{State graph of process scheduling status}\label{fig:process}
\end{figure}
Suppose \(Status(w)\in \set{free,  ready, running, blocked }\) is the scheduling
status of a process within an
operating system and the event alphabet is 
\(A_{sched} = \set{schedule,wake,preempt,run,block,free}\).
Scheduling status can be represented by a simple state graph
as in figure \ref{fig:process} or with the following sequential function.
\[\begin{array}{l} Status(\ess)=free\\
Status(w\xy a) = \begin{cases}
	ready &\text{if }Status(w)=free \text{ and }a=schedule\\
	&\text{or if }Status(w)=blocked \text{ and }a=wake\\
	&\text{or if }Status(w)=running \text{ and }a=preempt\\
	running &\text{if }Status(w)=ready \text{ and }a=run\\
	blocked &\text{if }Status(w)=running \text{ and }a=block\\
free &\text{if }Status(w)=running \text{ and }a=exit\\
	error&\text{otherwise}
\end{cases}\end{array}\]

	The only difference between this state machine and the one
	defined by the state graph is the unrecoverable error state
	on an unexpected event. We could just forbid those events
	with \(Status(w)\neq error\), but that constraint would then
	be on the wrong level because the schedule component does not control
	its inputs. 

To track which core a process is run on replace the single \(run\) event
with \((run,c)\) for \(c\in Cores\) where \(Cores\) is a set of processor core identifiers. 
To track the \emph{reason} a process is blocked, do the same thing with
\((block,r)\) and a set of \(Reasons\).
\[A_{sched} = \set{schedule,wake,preempt,(run,c)\ c\in Cores, (block,r)\ r\in Reasons,free}.\]
Then change the function definition.

\[ Status(w\xy a) = \begin{cases}
	ready &\text{if }Status(w)=free \text{ and }a=schedule\\
	&\text{or if }Status(w)=blocked \text{ and }a=wake\\
	&\text{or if }Status(w)=running \text{ and }a=preempt\\
	(running,c) &\text{if }Status(w)=ready \text{ and }a=(run,c)\\
	(blocked, r)&\text{if }Status(w)=running \text{ and }a=(block,r)\\
free &\text{if }Status(w)=running \text{ and }a=exit\\
	error&\text{otherwise}
\end{cases}\]
Even this change would make the graph in figure \ref{fig:process} impractical
to draw if either the set of core names or the set of reasons had more than 
a very few elements. 

	The specification does not satisfy:
	\[\text{If }Status(w)=ready\text{ then }(\eventually Status(w)=(running,c))\]
	for some \(c\in Cores\). The schedule component depends on the operating
	system for liveness as seen in section \ref{sec:multiple}. The techniques of that
	section would also permit interconnection  of other sequential functions that
	capture other process behavior.
	For example, \(Req(w)\in Requests\cup \set{null}\) might
	capture the current request to the operating system, if any.

\subsection{Clock pulse}\label{sec:clockpulse}
\begin{figure}[h]
\begin{center}  \includegraphics[width=0.5\textwidth]{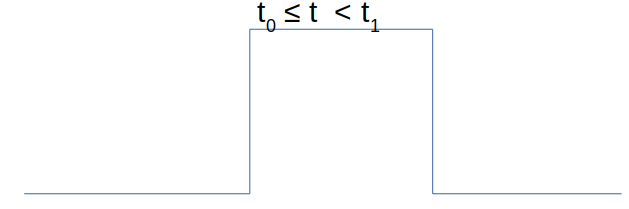} \end{center}

	\caption{Specifying a clock signal }
	\label{fig:dwell}
\end{figure}
Suppose the system events are binary \(n\)-tuples that represent
discrete time samples of input signals at a fixed frequency
so \((a)_i\) is the signal on wire
\(i\) carried by event (sample) \(a\). Because the frequency is fixed, counting events tracks the passage of time. Say \(C\) is a circuit with \(m\) output
signals if \(C(w)\in \set{0,1}^m\). Let \((C(w))_i\) be the \(i^{th}\) element
of that output tuple. Suppose output \(k\)
is supposed to be a clock signal with a cycle constrained by
time parameters \(t_0,t_1\) where \(t_1> t_0\) and the signal is 
	supposed to stay stable \emph{until} at least \(t_0\) time units
	and then switch levels before \(t_1\) time units have passed. 
See figure \ref{fig:dwell}. 
 What happens to the other output signals or what
effects the input signals have on them is not specified, yet. 
Define \(Stable(w,k, b)\) to count the time duration
that output \(k\) has been stable at value \(b\in \set{0,1}\).
\[\begin{array}{l}
	Stable(\ess,k,b)=0\\
		Stable(w\xy a,k,b) = ((C(w\xy a))_k = b)*(1+Stable(w,k,b))
	\end{array}\]
	Then the desired circuit property could be written:
	\[\text{If }C(w)_k=b\text{ then }C(w)_k=b \until Stable(w,k,b)\geq t_0\]
	and
	\[Stable(w,k,b) < t_1\]

	or we could replace the statement using \(\until\) with 
	\[\text{If }C(w)_k \neq C(w\xy a)_k\text{ then }Stable(w,k,b)\geq t_0\]
	
\subsection{Multiple processes\label{sec:multiple}}.

\newcommand{\dlta}[1]{{#1}^\mathsf{T}}
Suppose the schedule state systems from section \ref{sec:process}  are components of an operating system or network system
with its own event alphabet and there is set \(ProcessIds\) of process identifiers. For  \(p\in ProcessIds\), let:
\[ Pstatus(w,p) = Status(u_p(w))\]
where \(u_p\) is a \emph{connector} sequential function that translates sequences
of the enclosing system into sequences over the component alphabet 
\(A_{sched}\). 
When the system is
in the state determined by \(w\), schedule component \(p\) is in
the state determined by \(u_p(w)\).
The constraint \(Pstatus(w,p)\neq error\) requires the connector functions
to not generate events that the schedulers don't expect.

Connector functions are defined recursively:
\[ u_p(\ess)= \ess, \text{ and }  u_p(w\xy a) = u_p(w)\concat G_p(w,a).\]
The equation \(u_p(\ess)=\ess\) puts the process 
components in their initial states
when the operating system is in its initial state. The initial value can 
be any constant sequence over the component alphabet. The
map \(G_p\) is called the tail function and 
\(G_p(w,a)\) is what happens to the component \(p\) when the system advances
by an \(a\) event from the system state determined by \(w\). 
If \(G_p(w,a)=\ess\), component \(p\) remains in the same state. 
For
this simple example suppose each process component only advances \(0\) or \(1\) steps
on each step of the enclosing system and write \(a\) for \(\seq{a}\) if
it is clear from the context that the single element sequence being used.

\begin{figure}[h]
	\begin{center}
	\includegraphics[width=0.4\textwidth]{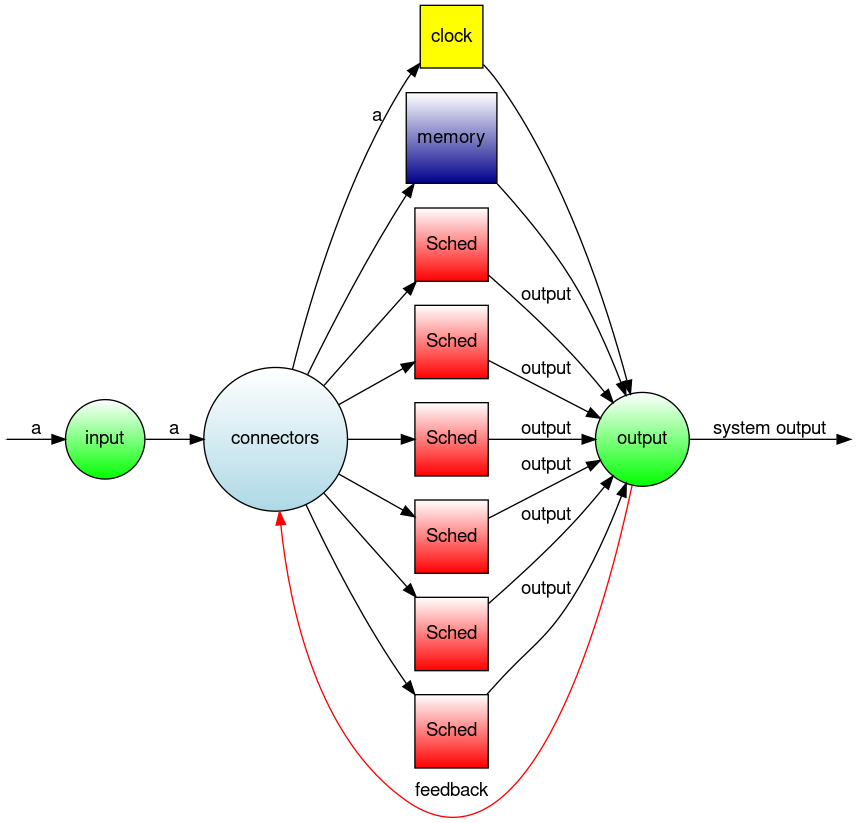} 
	\caption{Multiple process state machines, interconnected}\label{fig:mprocess}
	\end{center}
\end{figure}

There can be many different kinds of components in an interconnected system
and figure \ref{fig:mprocess} is a partial illustration of the operating system
with components. For the \(Req\) sequential function of section \ref{sec:process} there might
be connector maps \(V_p\) for each \(p\in Processes\).  
\[\begin{array}{l}G_p(w,a)={(block,r)}\text{ only if }Req(V_p(w))=r \text{ and}\\
\text{  for some }c,\  Sched(u_p(w))=(running,c)\end{array}\]

A requirement that no core is asked to run more
than one process in any state
constrains \(u_p\). 
\[\begin{array}{l}\text{For each }c\in Cores \text{ there is at most one }p\in ProcessIds,\\
\text{ so that } Pstatus(w,p)=(running,c)\end{array}\]
This property could be made more direct with:
\[\begin{array}{l}
	\text{If }G_p(w,a) = {(run,c)}\text{ then }Pstatus(w,p)=ready\\
	\text{    and }\Sigma_{p}(Pstatus(w,p)=(running,c)) = 0\\
	\text{    and }\Sigma_p(G_p(w,a)={(run,c))} = 1
\end{array}\]

%Let's add an event \(step\) to the alphabet of processes to indicate
%some internal computation.
%\[\begin{array}{l}
%	\text{If }g_p(w,a) = step\text{ then }Pstatus(w,p)=(running,c)\\
%	\text{ for some }c\in Cores\\
%\end{array}\]

Now it makes sense to require scheduling be live.
\[\begin{array}{l}
	\text{If }Pstatus(w,p)=ready\text{ then }\\
(\eventually Pstatus(w,p)=(running,c)) \text{ for some }c\end{array}\]
This property could be a useful
first approximation for example to show some network 
algorithm eventually gets to the right place.
On the other hand, general purpose
time-shared operating system generally cannot
guarantee this property.  We'd need
at least an admissions policy to limit the number of active processes
and additional controls on memory and other resources. A crude
requirement 
to keep the number of active processes below the number of cores.
\[\Sigma_p (Pstatus(w,p)\neq free) \leq |Cores|\]
would be interesting only for very simple real-time systems. Attempting
to specify how a general purpose operating system could guarantee
liveness or
at least identify the failure conditions that would break the 
guarantee, is a good problem. More realistically we could define a 
boolean sequence function to test whether timing specifications have
been met,
Then qualifying any propositions about the system
with with this test
limits those propositions to states where the timing specification
has been met.

\subsection{Real-time process constraints}\label{sec:realtime}
The enclosing system can contain additional components along with 
processes. One component could be a real-time clock, which may not correspond
to an actual device but could just be an ideal clock that somehow extracts
time from events at the enclosing system level.
\begin{equation}\label{eq:clock} Clock(w\xy a) >  Clock(w)\end{equation}
In a deterministic event driven system, the passage of real-time must be
represented by events since nothing else changes states but not 
all components need to change state as time advances.
Which events cause time to pass
can differ by what physical system is 
being represented. For the circuit above described by figure \ref{fig:dwell}, events are real-time samples of
signal values so every event marks passage of one unit of time.
In the operating system or network context, the passage
of time might be marked by 
clock ticks from an oscillator, or clock interrupts,
or some kind of input value like a time
received from a GPS satellite or a vector of samples of signals on all 
input wires of the top level system during a discrete moment of time. It
is also possible to take a real valued time variable \(t\) and just
require that \begin{equation}\label{eq:clock2} Clock(w) \sim t\ .\end{equation} 
That choice doesn't have to be made at this point. 
Let
\begin{equation}\label{eq:clock3}
Elapsed(w,a) = Clock(w\xy a)-Clock(w).\end{equation}
Now it is possible to track how long a process has been runnable, waiting
to run.
Define \(Delay(w,p)\) as follows:
\
\begin{eqnarray*}Delay(\ess,p)=0\\ Delay(w\xy a,p) = \begin{cases}
	Delay(w,p)+Elapsed(w,a)&\text{if }Pstatus(w\xy a,p) = ready\\
	&\text{and }Pstatus(w,p)=ready\\
	0&\text{otherwise}\end{cases}\end{eqnarray*}
	\(Clock\) depends on \(w\) the sequence driving the operating system 
	or network directly, while \(Pstatus(w,p)\) depends on \(u_p(w)\).

	Define \(RTSched(w,x)\) for \(x>0\)
	by \(RTSched(\ess,x)=1\)
	and
	\[RTSched(w\xy a) = \begin{cases}
		0&\text{if for some }p,\ Delay(w\xy a,p) >x\\
		RTSched(w,x)&\text{otherwise}
	\end{cases}\]

	To require that every scheduled process must \emph{eventually} run:
	\[\text{for some }x,\ RTSched(w,x)\]

It is certainly possible to imagine the bound being state dependent, so \[Delay(w,p) \leq Bound(w).\]
	\(Bound\) might depend on how many processes are waiting to run and
	what resources they need and their priorities. 

	The constraint using the \(\eventually\) operator
is not the same as any of these,
because the eventually operator counts each event as one
time unit and keeps the count invisible.

\section{A Message queue} \label{sec:abadi}

\begin{figure}[ht]
	\begin{center}
	\includegraphics[width=0.5\textwidth]{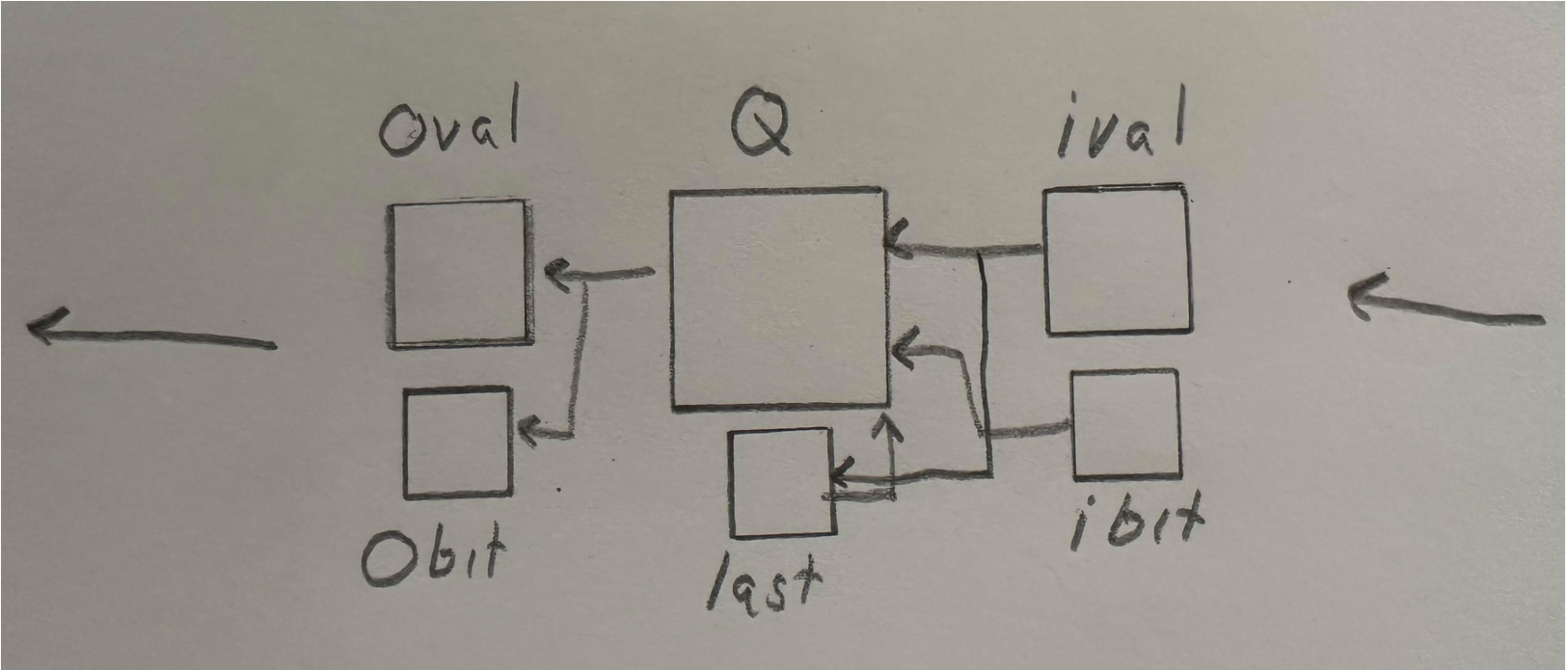}
	\caption{The queue from Abadi/Lamport}\label{fig:abadiq}
	\end{center}
\end{figure}

A paper by Abadi and Lamport \cite{abadilamport} includes an example of a somewhat
peculiar buffered queue.
The queue is first specified in 
a way that
makes it unreliable (it can lose input data) and then it is fixed 
by adding real-time 
constraints. I've tried to stay close to the approach in the Abadi and Lamport
paper which they summarize as follows:

\begin{quote}\textit{
	The
	interface consists of two pairs of ``wires'', each pair consisting of a \emph{val} wire that
	holds a message and a boolean-valued bit wire. A message \emph{m} is sent over a pair of
	wires by setting the \emph{val} wire to \emph{m} and complementing the \emph{bit} wire. The receiver
detects the presence of a new message by observing that the bit wire has changed
	value. Input to the queue arrives on the wire pair (\emph{ival, ibit}),
	and output is sent
	on the wire pair (\emph{oval, obit}). There is no acknowledgment protocol, so inputs are
	lost if they arrive faster than the queue processes them. (Because of the way \emph{ibit}
is used, inputs are lost in pairs.) The property guaranteed by this lossy queue is
that the sequence of output messages is a subsequence of the sequence of input
	messages\footnote{This is not correct. VY}. [Below] we add timing constraints to rule out the possibility of
	lost messages.} (page 1545)
\end{quote}

\begin{figure}[ht] 
	\begin{center}
	\includegraphics[width=0.7\textwidth]{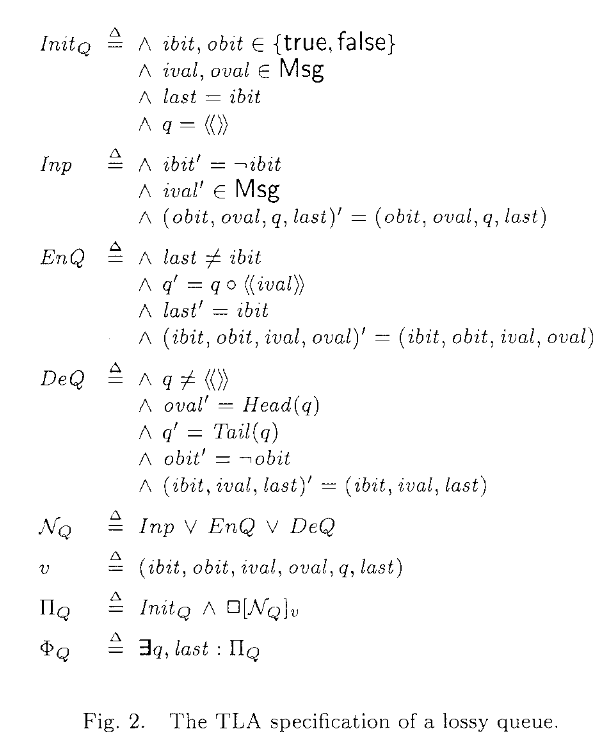}
	\caption{Abadi and Lamport's queue specification (untimed)}
	\label{fig:abadi}
	\end{center}
\end{figure}

The core of the pre-realtime 
Abadi and Lamport specification is in figure \ref{fig:abadi}
but the reader here doesn't need to examine it to understand this specification.

The motivating example might be a A/D device or simple input device that has a single
datum (message) input buffer, a program or device that is responsible for collecting and queuing the messages, and a third component that dequeues messages and makes them available to a consumer in the output buffer. 
The queue is lossy because is is possible that 
messages that are stored to the input buffer can be overwritten by a second
\(input\) event before they are moved into the internal queue.

\newcommand{\Msgs}{\mathit{Msgs}}
From the outside, ignoring the components for the moment,
suppose there is an alphabet \(A_{queue}\) which contains 
events including \(deq\) and \((input,m)\) for every message \(m\in \Msgs\)
and other events which can be left unspecified. The output is a pair
of the message at the head of the queue and a bit that alternates on each
message. 
\[(m, b) \Leftarrow \colorbox{blue!40}{Q} \Leftarrow  enq,(input,m),\dots\]
The output message is \(nullm\) if the queue was empty on the last \(deq\).

The queue
is constructed from the store, toggle, and queue components from
specification \ref{spec:simple} above. 
\begin{spec}
	\(\Msgs\)  is a set of messages.\\
	\(Inval\) and \(Outval\) are stores over \(\Msgs\) with initial 
	value \(nullm\);\\
	\(Inb\), \(Outb\), and \(Last\) are toggles with the same initial value.\\
	\(Q\) is a transparent queue over \(\Msgs\);\\
\end{spec}

A \((input,m)\) event stores message \(m\) in the input buffer and toggles
the input toggle. A \(deq\) event causes the internal queue to move 
the head of the queue to the output buffer and toggles the
output toggle. An internal \(enq\) event copies the message in
the input 
buffer to the internal queue and flips the \(Last\) toggle bit. 
The intent of the protocol with the toggles is that  the \(Inb\)
and \(Last\) toggles are not equal if and only if
there is a live message in 
\(Inval\) tht has not yet been enqued in the internal queue.

The connector maps are \(w_1\) for \(Inb\) and \(Inval\), 
\(w_2\) for the toggle \(Last\), \(w_3\) for the internal queue \(Q\),
and \(w_4\) for the output bit toggle and output buffer. Say
\(w_i(w\xy a)\) \emph{adds} an event \(a\) if and only if
\(w_i(w\xy a) = w_i(w)\xy a\). Say
\(w_i(w\xy a)\) \emph{adds} nothing if and only if
\(w_i(w\xy a) = w_i(w)\).
The specification
for \(w_3\) is different from the others because we don't want to fully
specify when \(w_3(w\xy a)\) adds an \(enq\) event. In the Abadi and Lamport
specification this is a nondeterministic choice, here it depends on 
the unspecified variations among the components, the other input events, 
and possibly other factors not specified. Section \ref{sec:rt} adds real-time
constraints. 

\begin{spec}
\[\begin{array}{l}
	RawDevice(w) = (Inval(w_1),Inb(w_1), Last(w_2), Q(w_3),Outval(w_4),Outb(w_4)) \\
		\\
	w_i(\ess)=\ess\\
	w_1(w\xy a) = \begin{cases}
		w_1(w)\xy a&\text{if }a=(input,m)\text{ for some }m\in Msgs\\
	w_1(w)&\text{otherwise}\end{cases}\\
	w_2(w\xy a) = \begin{cases} 
		w_2(w)\xy a&\text{if }w_3(w\xy a)=w_3(w)\xy (enq,m)\\
		        \text{       for some } m\in Msgs\\
	w_2(w)&\mbox{otherwise}\end{cases}\\
	w_4(w\xy a) = \begin{cases}
		head(Q(q(w)))&\text{if }a=deq\\
		&\text{and } Q(w_3(w))\neq \ess\\
	w_4(w)&\text{otherwise}\end{cases}\\
	\\
	w_3(w\xy a) \text{ adds }deq, (enq,Inval(w_1(w)), \text{ or }nothing.\\
	w_3(w\xy a)\text{ adds }deq\text{ if and only if }a=deq\\
	w_3(w\xy a)\text{ adds }(enq,Inval(w_1(w)))\text{ only if } a\neq deq\\
		\text{       and }Last(w_2(w))\neq Inb(w_1(w))\\
	w_3(w\xy a)\text{ adds nothing if and only if }a\neq deq\\
		 \text{      and }Last(w_2(w)) = Inb(w_1(w))\\
\end{array}\]

%\end{array}\]

	\end{spec}

	The queue device itself then just selects from the tuple
	value of \(RawDevice\).

	\begin{spec}
	\[Queue(w) = ((RawDevice(w))_5, (RawDevice(w))_6)\]
	\end{spec}

		\subsection{Real-time\label{sec:rt}}
To patch up the poor design of the protocol
in this queue, Abadi and Lamport propose to put
timing constraints on input and enq operations. The basic idea here
is that when an \((input,m)\) event happens, a timer starts and the matching
\(enq\) must happen before \(t_0\) seconds pass, but  the next
input event cannot happen until more than \(t_0\) seconds have passed.
It's pretty obvious that these timing constraints fix the problem
with dropped messages, but see section \ref{sec:correctness}. 

Let's use the \(Clock\) described in equation \ref{eq:clock} and 
\ref{eq:clock2}  and \(Elapsed\) from equation \ref{eq:clock3} above. 

\[\begin{array}{l} (oval, obit) \Leftarrow \colorbox{blue!40}{Q} \Leftarrow  enq,(input,m)\\
time  \Leftarrow \colorbox{yellow!40}{Clock} \Leftarrow  A_{queue}\end{array}
	\]
\(SinceI\) tracks  the time that has passed since the last \(input\) event. In the initial
state, it is set to \(t_0\) so that the initial \(input\) can
happen at any time without flagging an error.

\[\begin{array}{l}
SinceI(\ess)= t_0\\
SinceI(w\xy a)= \begin{cases} 
	0 &\text{if }	a=(input,m)\\
	&\text{ for some }m\in \Msgs\\
SinceI(w)+Elapsed(w,a)&\text{otherwise}\end{cases}
\end{array}	\]
There cannot be an \((input,m)\) event without triggering
a timing error \emph{until} \(SinceI(w) > t_0\). But also
once there is an input event, 
there must be an \(enq\) event \emph{before} \(SinceI(w) \geq t_0\).

To make the specification more compact:
\(WaitEnq\) indicates if there is a live message in the inbuffer according
to the protocol. 
\[	WaitEnq(w) = \begin{cases} 1&\mbox{ if }Last(w_2(w))) \neq Inb(w_1(w)))\\
0&\text{otherwise}\end{cases}\]

Putting these together the timing requirement is:
	\begin{equation}\begin{array}{l} Tspec(\ess) = 1\\
 Tspec(w\xy a) = \begin{cases}
	 0 &\text{if }WaitEnq(w\xy a) \text{ and }SinceI(w\xy a) \geq t_0\\
	 &\text{or if }a=(input, m)\\&\text{ for some }m\in \Msgs\\
		&\text{ and } SinceI \leq t_0\\
	Tspec(w)&\text{otherwise}
 \end{cases}\end{array}
\end{equation}
	Once \(Tspec(w)=0\) it stays that way, because a single timing error
	is not recoverable for this queue. 

	The timing specification is not requiring that events follow
	the specification which could be done
	with \(\always Tspec(w)=1\). Instead
	the specification allows us to distinguish states that are
	reached without violations of the specification. Those reached
	otherwise have unspecified results.

	For comparison, see page 1553 of \cite{abadilamport} which is 
	similar in intent.

\subsection{Correctness}\label{sec:correctness}

	We want to prove that if \(Tspec(w)=1\) then the queue has 
	not dropped or reordered or invented any queue elements.
	Abadi and Lamport characterize this property
	as something like, the sequence 
	of input values is a prefix of the sequence of output values. We're
	going to prove something stronger but first, let's define what those
	two sequences are.

Define: \(Q_I(\ess)=\ess\) and 
	\[ Q_I(w\xy a) = \begin{cases}
		Q_I(w)\xy m&\text{if }a=(input, m)\text{ for some }m\in \Msgs\\
	Q_I(w)&\text{otherwise}\end{cases}\]
This is the abstract queue of inputs. 
Define: \(Q_O(\ess)=\ess\) and 
	\[ Q_O(w\xy a) = \begin{cases}
		Q_O(w)\xy m&\text{if }w_3(w\xy a)=w_3(w)\xy (deq,m)\text{ for some }m\in \Msgs\\
	Q_O(w)&\text{otherwise}\end{cases}\]

	This is the queue of messages that has passed through the device to
	the output buffer. 
The queues are similar to the \emph{history variables} of Abadi and Lamport.
	The theorem (it's really too simple to be called
	a theorem) to prove is stronger than what Abadi and 
	Lamport suggest. 
	\begin{thm} \label{thmwait}
		Assuming \(Tspec(w)\)
		\[\text{If }WaitEnq(w)=0\text{ then } Q_I(w) = Q_O(w)\concat Q(w_3(w))\]
		\[\text{If }WaitEnq(w)=1\text{ then } Q_I(w) = Q_O(w)\concat Q(w_3(w))\xy Inval(in(w)) \]

	\end{thm} 
	In either case, \(Q_O(w)\) is a prefix of \(Q_I(w)\).
	The difference is that when \(WaitEnq(w)=1\) the value in the
	input buffer is the last item that has shown up on the input queue,
	otherwise the input buffer can be ignored. This theorem also shows
	that the protocol works as long at \(Tspec(w)\) remains true.

	The proof of \ref{thmwait} is by induction on the event sequence.
	The proof here is in too much detail, just to show it can be
	done completely.
	Initially, when \(w=\ess\), all the queues are empty and
	\(WaitEnq(\ess)=0\) so there
	is nothing to prove. Suppose the theorem is true for \(w\)
	and consider \(w\xy a\). We assume \(Tspec(w\xy a)\) otherwise
	there is nothing to prove. There are four cases depending
	on the four possible values of \((WaitEnq(w),WaitEng(w\xy a))\).
	\begin{enumerate}
		\item case: \[WaitEnq(w)=0 \text{ and }WaitEnq(w\xy a)=1\]
	By the definition of \(WaitEnq\), one of the 
			two toggles has to change thanks to \(a\). Because \(WaitEnq(w)=0)\), 
			\(w_3(w\xy a)\) 
			cannot add a \((deq,m)\) event
			that would change the \(Last\) toggle. 
			So the alternative must be that \(a=(input,m)\)
			for some \(m\).
	\(m\) and \(Inb\) changes.
	By the inductive assumption and because \(WaitEnq(w)=0\)
	\(Q_I(w) = Q_O(w)\concat Q(w_3)\).
	Because \(a= (input,m)\), 
	\(Q_I(w\xy a) = Q_I(w)\xy m\).
			By the definition of \(w_1\), it must be
			that \(Inval(w\xy a)=m\)
			We know \(a\neq deq\) so \(w_3(w\xy a)\)
			cannot add \(deq\) -- so 
	\(Q(w_3(w)) = Q(w_3(w\xy a))\). It follows that
			\(Q_I(w\xy a) = Q_O(w\xy a)\concat Q(w_3(w\xy a)\xy Inval(w_1(w\xy a)\). 
		\item case: \[WaitEnq(w)=0 \text{ and }WaitEnq(w\xy a)=0\]
	In this case \(a\neq (input,m)\) because otherwise
	\(WaitEnq(w\xy a) = 1\) unless the \(Last\) toggle flips too.
			But \(Last\) cannot change because \(WaitEnq(w)=0\)
			means that \(w_3(w\xy a)\) does not add
			an \((enq,m)\) which is required to toggle \(Last\).
			Because \(a\neq (input,m)\) it
			follows that \(Q_I(w\xy a))=Q_I(w))\). 
	If \(a = deq\) then if \(Q(w_3(w))=\ess\) nothing happens to
			\(Q_I\),
			\(Q_O\) or \(Q\).
			If \(Q(w_3(w))\neq \ess\) and \(a=deq\), then 
			\(w_3(w\xy a) = w_3(w)\xy deq\) so
			\(Q(w_3(w\xy a) = Tail(Q(w_3(w)))\)
	and \(Q_O(w\xy a) = Q_O(w)\xy Head(Q(w_3(w)))\) but
	\[Q_O(w)\xy Head(Q(w_3(w)) \concat Tail(Q(w_3(w))\]
	must be the same as 
	\[Q_O(w) \concat Q(w_3(w)).\]
			Otherwise \(w_3(w\xy a)\) adds
			\(\ess\) and nothing changes. 
	That takes care of this case.

		\item case: \[WaitEnq(w)=1 \text{ and }WaitEnq(w\xy a)=1\]
			Because \(WaitEnq\) remains true, \(w_3(w\xy a)\) does not add \((enq,m)\)
	for any \(m\) so no new elements are added to \(Q(w_3(w\xy a))\).
			Because we assume \(Tspec(w\xy a)\), \(WaitEnq(w)*SinceI(w) < t_0\) so  \(a\) is not an \((input,m)\) for any \(m\) 
			and thus \(w_1(w\xy a)= w_1(w)\)
			and \(Inval(inv(w\xy a))=Inval(inv(w))\).
	If \(a\neq deq\), it must
			be that \(w_3(w\xy a)\) adds \(\ess\) so \(Q(w_3(w\xy a))= Q(w_3(w))\) 
	 and similarly \(Q_O(w\xy a)=Q_O(w)\) because
			\(w_4(w\xy a)i = w_4(w)\). If \(a=deq\) then  
			then \(Q(w_3(w\xy a)) = Tail(Q(w_3(w)))\) and
			\(Q_O(w\xy a)= Q_O(w)\xy Head(Q(w_3(w)))\) so the constructed queue is unchanged. 

	For any other \(a\) nothing changes in the outputs of any components
	 and we are done with this case.
	
		\item case: \[WaitEnq(w)=1 \text{ and }WaitEnq(w\xy a)=0\]
	 Since \(WaitEnq(w)=1\)
	 \[Q_I(w) = Q_O(w)\concat Q(w_3(w))\xy Inv(inv(w).\] Either
	 \(SinceI(w) < t_0\) or \(TSpec(w)=0\). We're assuming
	 the second is false, so by the same reasoning
	 \(a\neq (input,m)\) (or else \(Tspec(w\xy a)=0\)). That means
	 that 
	 \(w_3(w\xy a)\) adds \((enq,Inv(w_1(w)))\) to make \(WaitEnq(w\xy a)=0\).
	 As a result \[Q(w_3(w\xy a)) = Q(w_3(w))\xy Inv(w_1(w)).\]
	 And the rest follows.
	\end{enumerate}

\subsection{Some notes}
The original specification depends on an unbounded queue, but in this case imposing a 
bound \(k\) on the queue could be accomplished by replacing the transparent
queue with a bounded queue of length \(k\) which 
would make an enq operation when the queue is full just discard the message in the in buffer but would still toggle the value in \(Last\). 
Given a timing specification for \(deq\) we could show that for some 
\(k\) the queue would be long enough. An interesting way to specify
timing would be to require that deq operations happen every \(t_1\)
seconds at the latest, but that once a deq operation happened, other 
deq's would happen every \(t_2\) for some \(t_2< t_1 \) until the queue
had been emptied. 

\section{Discussion}\label{sec:discussion}

When Krithivasan Ramamritham gave me temporal logic papers to read\cite{krithi}
at the start of my graduate studies in the 1980s, I had
just finished working on a commercial distributed operating system\cite{Borg} and was frustrated by how often 
developers found design errors
only during test after writing enormous quantities of code.
Temporal logic seemed like a solution, but I was not unique in concluding
is was not a satisfactory answer after attempting to apply it to
real systems. Lamport and Abadi argued (in 1994) that \emph{"It has long
been known that the methods originally developed for shared memory
multiprocessing apply equally well to distributed systems"} (p. 1543)
but it's not clear that they do apply well at all. 
Much of the complexity
of the Abadi and Lamport paper comes
from the assumption that there is a 
``fair scheduler'' somewhere deciding which nondeterministic action to 
activate -- for example, to advance the time variable.

This view is reflected in the 
use of 
infinite sequences of assignment maps (maps assigning values
to formal variables ) as formal semantics.
Infinite sequences
of assignment maps that must be composed by interleaving and where
semantics depends a scheduler that exists outside the 
quantifiers  inflexibility of the temporal quantifiers, the
\emph{ad hoc} nature of the formal semantics were among the issues that made it difficult to get a handle

Some temporal logic variants use infinite sequences of 
\emph{assignment} maps (maps from propositions or variables to
values) as formal semantics. 
modal and temporal logics can define state machines or use state
machines as semantic models.
Kripke's initial modal logic paper\cite{Kripke} makes this clear
although he did not use the terminolgy of state machines.
Finite state machines, of a kind, can be constructed from logical
assertions to serve
as models for temporal logic as well\cite{Clarke87,Clarke}:
\begin{quote}
Temporal logic formulae are interpreted over a given finite
	state graph, also called a (Kripke) structure, \(M\) comprised of
	a set \(S\) of states, a total binary transition relation \(R \subseteq S \times S\),
	and a labelling \(L\) of states with atomic facts (propositions)
true there. -- Edmond Clarke \cite{clarketuring}\end{quote}
These are 
non-deterministic machines of a sort with unlabeled state
transitions which may be controlled
by logical predicates. States are
assignment maps assigning values to formal symbols,
and composition is by interleaving and essentially taking the union of the
assigment maps assuming the domains are disjoint. The view is still
derived from concurrent processes in a timeshared single processor
computer system and there is clear model of composition. 

Methods used here and in \cite{yodaikenlarge}
were present in a 1991 paper \cite{yodaikenipl}, which included Moore machines,
	automata products for composition and concurrency and
	even a \emph{modal} primitive recursive specification 
	language. But that
	work was in the tradition of computational formal logic
	and this is not. 
	When computer scientists started to try to develop mathematical
	methods of specifying programs, they often
	adopted methods from formal
	logic and other axiomatic methods
	perhaps because of the role of
	formal languages in programming languages and pattern matching
	and because of the influence of metamathematics on computability
	studies. Whatever the merits of that approach,
	discarding the apparatus of formal logic 
	has allowed solution of a number of problems that blocked
	progress in the earlier method  and greatly simplified
	specifications. 

	The quote from Von Neumann that starts this paper is from an
	enormously perceptive paper (not unusual for the author)
	that among other things previews 
	algorithmic complexity theory. The full paragraph in which 
	that quote is found has an interesting critique which I have tried to learn from as I returned to this work after a 
	long absence. 

	\begin{quote}
		We are very far from possessing a theory of automata which deserves that name, that is, a properly mathematical-logical theory. There exists today a very elaborate system of formal logic, and, specifically, of logic as applied to mathematics. This is a discipline with many good sides, but also with certain serious weaknesses. This is not the occasion to enlarge upon the good sides, which I have certainly no intention to belittle. About the inadequacies, however, this may be said: Everybody who has worked in formal logic will confirm that it is one of the technically most refractory parts of mathematics.
The reason for this is that it deals with rigid, all-or-none concepts, and has very little contact with the continuous concept of the real or of the complex number, that is, with mathematical analysis.
Yet analysis is the technically most successful and best-elaborated part of mathematics. Thus formal logic is, by the nature of its approach, cut off from the best cultivated portions of mathematics, and forced onto the most difficult part of the mathematical terrain, into combinatorics.
	\end{quote}

%\bibliography{all}
%\bibliographystyle{plain}
\printbibliography

\appendix
\section{Appendix: Sequence utilities\label{appendix:sequences}}

	 \[w\concat \ess = w,\ w\concat (z\xy a) = (w\concat z)\xy a)\]
	 \[length(\ess) = 0,\ length(w\xy a) = 1+ length(w)\]
	 \[prefixes(\ess)= \set{\ess},\ prefixes(w\xy a) = prefixes(w)\cup\set{w\xy a}\]
	 \[rappend(\ess,a) = a\xy \ess, rappend(w\xy b,a)=rappend(w,a)\xy b)\]
	 \[Head(\ess)=\ess, Head(rappend(w,a)) = a\]
	 \[Tail(\ess)=\ess, Tail(rappend(w,a)) = w\]
\end{document}